\documentclass[a4paper]{article}

\usepackage{INTERSPEECH2021}

\usepackage{float}
\usepackage{times}
\usepackage{latexsym}
\usepackage{multirow}

\usepackage[utf8]{inputenc}
\usepackage{soulutf8}
\usepackage{todonotes}
\usepackage{placeins}
\usepackage{tikz}
\usepackage{svg}
\usepackage{hyperref}

\newcommand\citep{\cite}

\usepackage{url}

\usepackage{mathtools,calc}

\makeatletter
\newcommand*\concat
  {\mathbin{\mathmakebox[\widthof{${+}\m@th$}]{+\hskip-1emplus1fil+}}}

\makeatother

\title{TTS-Portuguese Corpus: a corpus for speech synthesis in Brazilian Portuguese}

\name{Edresson Casanova$^1$,  Arnaldo Candido Junior$^2$, Christopher Shulby$^3$, Frederico Santos de Oliveira$^4$, Jo\~ao Paulo Teixeira$^5$, Moacir Antonelli Ponti$^1$, Sandra Maria Aluisio$^1$}
\address{\normalsize
    $^1$ Instituto de Ciências Matemáticas e de Computação, University of S\~ao Paulo, S\~ao Carlos/SP, Brazil \\
    $^2$ Federal University of Technology – Paraná, Medianeira, PR, Brazil \\
    $^3$ DefinedCrowd Corp., Seattle, WA, USA \\
    $^4$ Federal University of Mato Grosso, Cuiab\'a - MT, Brazil \\
    $^5$ Research Center in Digitalization and Intelligent Robotics (CEDRI), Bragança, Portugal 
  }
\email{edresson@usp.br}

\begin{document}
\maketitle

\begin{abstract}
Speech provides a natural way for human-computer interaction. In particular, speech synthesis systems are popular in different applications, such as personal assistants, GPS applications, screen readers and accessibility tools. However, not all languages are on the same level when in terms of resources and systems for speech synthesis. This work consists of creating publicly available resources for Brazilian Portuguese in the form of a novel dataset along with deep learning models for end-to-end speech synthesis. Such dataset has 10.5 hours from a single speaker, from which a Tacotron 2 model with the RTISI-LA vocoder presented the best performance, achieving a 4.03 MOS value. The obtained results are comparable to related works covering English language and the state-of-the-art in Portuguese.
\end{abstract}


\section{Introduction}

Speech synthesis systems have received a lot of attention in recent years due to the great advance provided by the use of deep learning, which allowed the popularization of virtual assistants, such as Amazon Alexa \citep{purington2017alexa}, Google Home \citep{dempsey2017teardown} and Apple Siri \citep{siri}. 

\color{black}

According to \cite{dctts} traditional Speech Synthesis systems are not easy to develop, because these are typically composed of many specific modules, such as, a text analyzer, a grapheme-to-phoneme converter, a duration estimator and an acoustic model. 
In summary, given an input text, the text analyzer module converts dates, currency symbols, abbreviations, acronyms and numbers into their standard formats to be pronounced or read by the system, i.e. carries out the text normalization and  tackles problems like homographs, then with the normalized text the phonetic analyzer converts the grapheme into phonemes. In turn, the duration estimator estimates the duration of each phoneme. Finally, the acoustic model receives the phoneme representation sequence, the prosodic information about phoneme segments' length, the F0 contour and computes the speech signal \citep{ze2013statistical, teixeira2003prediction}. Several acoustic models have been proposed such as the classical formant model \citep{klatt1980software}, Linear Prediction Coefficients (LPC) model, the Pitch Synchronous Overlap and Add (PSOLA) models \citep{charpentier1986diphone} widely used in TTS engines like Microsoft Speech API. In addition, Hidden Markov Model (HMM) based synthesis is still a topic of research \citep{tokuda2000speech,braude2013template,aroon2015statistical}, as well as a variety of Unit Selection Models \citep{wang2011automatic,siddhi2017survey}. 
\color{black}

Deep Learning \citep{deeplrgoodfellow} allows to integrate all processing steps into a single model and connect them directly from the input text to the synthesized audio output, which is referred to as end-to-end learning. While neural models are sometimes criticized as difficult to interpret, several end-to-end trained speech synthesis systems \citep{kyle2017char2wav,tacotron,tacotron2,dctts,deepvoice3, kim2020glow, valle2020flowtron} were shown to be able to estimate spectrograms from text inputs with promising performances. Due to the sequential characteristic of text and audio data, recurrent units were the standard building blocks for speech synthesis such as in Tacotron 1 and 2 \citep{tacotron,tacotron2}. In addition, convolutional layers showed good performance while reducing computational costs as implemented in DeepVoice3 and Deep Convolutional Text To Speech (DCTTS) methods~\citep{dctts,deepvoice3}.
\color{black}

Models based on Deep Learning require a greater amount of data for training, therefore, languages with low available resources are impaired. For this reason, most current TTS models are designed for the English language \citep{tacotron2,deepvoice3, valle2020flowtron, kim2020glow}, which is a language with many open resources. In this work we propose to solve this problem for the Portuguese language. Although there are some public datasets of synthesis for European Portuguese \citep{teixeira2001phonetic}, due to the small amount of speech, approximately 100 minutes, makes the training of models based on Deep Learning unfeasible. In addition, simultaneously with this work, two datasets for automatic speech recognition for Portuguese, with good quality, were released. The CETUC \citep{alencar2008lsf} dataset, which was made publicly available by \citep{quintanilha2020open}, has approximately 145 hours of 100 speakers. In this dataset, each speaker uttered a thousand phonetically balanced sentences extracted from journalistic texts; on average each speaker spoke 1.45 hours. The Multilingual LibriSpeech (MLS) \citep{pratap2020mls} dataset is derived from LibriVox audiobooks and consists of speech in 8 languages including Portuguese. For Portuguese, the authors provided approximately 130 hours of 54 speakers,  an average of 2.40 hours of speech per speaker. Although the quality of both datasets is good, both were made available with a sampling rate of 16Khz and have no punctuation in their texts, making it difficult to apply them for speech synthesis. In addition, the amount of speech per speaker in both datasets is low, thus making it difficult to derive a single-speaker dataset with a large vocabulary for single-speaker speech synthesis. For example, the LJ Speech \citep{ito2017lj} dataset, which is derived from audiobooks and is one of the most popular open datasets for single-speaker speech synthesis in English, has approximately 24 hours of speech.

\color{black}

In this article, we compare models of TTS available in the literature for a language with low available resources for speech synthesis. The experiments were carried out in Brazilian Portuguese and based on a single-speaker TTS. For this, we created a new public dataset, including 10.5 hours of speech. Our contributions are twofold (i) a new publicly available dataset with more than 10 hours of speech recorded by a native speaker of Brazilian Portuguese; (ii) an experimental analysis comparing two publicly available TTS models in Portuguese language.
In addition, our results and discussions shed light on the matter of training end-to-end methods for a non-English language, in particular Portuguese, and the first public dataset and trained model for this language are made available.

This work is organized as follows. Section \ref{sec:related} presents related work on speech synthesis. Section \ref{sec:method:base} describes our novel audio dataset. Section \ref{sec:method}  details the models and experiments performed. Section \ref{sec:results} compares and discusses the results. Finally, Section \ref{sec:conc} presents conclusions of this work and future work.

\section{Speech Synthesis Approaches} \label{sec:related}

With the advent of deep learning, speech synthesis systems have evolved greatly, and are still being intensively studied. Models based on Recurrent Neural Networks such as Tacotron \citep{tacotron}, Tacotron 2 \citep{tacotron2}, Deep Voice 1 \citep{deepvoice} and Deep Voice 2 \citep{deepvoice2} have gained prominence, but as these models use recurrent layers they have high computational costs. This has led to the development of fully convolutional models such as DCTTS \citep{dctts} and Deep Voice 3 \citep{deepvoice3}, which sought to reduce computational cost while maintaining good synthesis quality.

\cite{deepvoice3} proposed a fully convolutional model for speech synthesis and compared three different vocoders: Griffin-Lim \citep{griffin1984signal}, WORLD Vocoder \citep{world} and WaveNet \citep{WaveNet}. Their results indicated that WaveNet neural vocoder produced a more natural waveform synthesis. However, WORLD was recommended due to its better runtime, even though WaveNet had better quality. The authors further compared the proposed model (Deep Voice 3) with the Tacotron \citep{tacotron} and Deep Voice 2 \citep{deepvoice2} models.

\cite{dctts} proposed the DCTTS model, a fully convolutional model, consisting of two neural networks. The first, called Text2Mel (text to Mel spectrogram), which aims to  generate a Mel spectrogram from an input text and the second, Spectrogram Super-resolution Network (SSRN), which converts a Mel spectrogram to the STFT (Short-time Fourier Transform) spectrogram \citep{benesty2011speech}. DCTTS consists of only convolutional layers and uses dilated convolution \citep{yu2015multi,kalchbrenner2016neural} to take long, contextual information into account. DCTTS uses the vocoder RTISI-LA (Real-Time Iterative Spectrogram Inversion with Look-Ahead) \citep{zhu2007real}, which is an adaptation of the Griffin-Lim vocoder \citep{griffin1984signal}, which aims to increase the speed of the synthesis by slightly sacrificing the quality of the audio generated.

Tacotron 1 \citep{tacotron} proposes the use of a single trained end-to-end Deep neural network. The model includes an encoder and a decoder. It uses an attention mechanism \citep{bahdanau2014neural} and also includes a post-processing module. This model uses convolutional filters, skip connections \citep{srivastava2015highway}, and Gated Recurrent Units (GRUs) \citep{chung2014empirical} neurons. Tacotron also uses Griffin-Lim \citep{griffin1984signal} algorithm to convert the STFT spectrogram to the wave form.

Tacotron 2 \citep{tacotron2} combines Tacotron 1 with a modified WaveNet vocoder \citep{wavenetmodificado}. Tacotron 2 is composed of a recurrent network of prediction resources from sequence to sequence that maps the incorporation of characters in Mel spectrograms, followed by a modified WaveNet model acting as a vocoder to synthesize waveforms in the time domain from those spectrograms. They also demonstrated that the use of Mel spectrograms as the conditioning input for WaveNet, instead of linguistic characteristics, allows a significant reduction in the size of the WaveNet architecture. 
\\\\

\section{TTS-Portuguese Corpus}\label{sec:method:base}

\color{black}
Portuguese is a language with few publicly available resources for speech synthesis. In Brazilian Portuguese, as far as we know there is no public dataset with a large amount of speech and quality available for speech synthesis. Although there are some public speech datasets for European Portuguese, for example \cite{teixeira2001phonetic}, the work has a small amount of speech, approximately 100 minutes, which normally is not useful for training deep-learning models. On the other hand, \cite{ttspropor} explored the training of a model based on deep learning with an in-house dataset, called SSF1, which has approximately 14 hours of speech in European Portuguese.  Therefore, given the inexistence of an open dataset with a large amount of speech and quality for speech synthesis for Brazilian Portuguese, we propose the TTS-Portuguese Corpus. 

To create the TTS-Portuguese Corpus, public domain texts were used. Initially, seeking to reach a large vocabulary we extracted articles from the Highlights sections of Wikipedia for all knowledge areas. After this extraction, we separated the articles into sentences (considering textual punctuation) and randomly selected sentences from this corpus during the recording. In addition, we used 20 sets of phonetically balanced sentences, each set containing 10 sentences proposed by \cite{seara1994estudo}. Finally, in order to increase the number of questions and introduce more expressive speech, we extracted sentences from Chatterbot-corpus\footnote{https://github.com/gunthercox/chatterbot-corpus/}, a corpus originally created for the construction of chatbots. Therefore, we decided both to have a large vocabulary and also to bring words from different areas. In addition, to have an expressive speech representation with the use of questions and answers from a chatbot dataset.

The recording was made by a male native Brazilian Portuguese speaker, not professional, in quiet  environment but without acoustic isolation due to difficulties having access to studios. All the audios were recorded at a sampling frequency of 48 kHz and a 32-bit resolution.

\begin{figure*}[!htbp]\small
 \centering
    \includegraphics[width=0.4\textwidth]{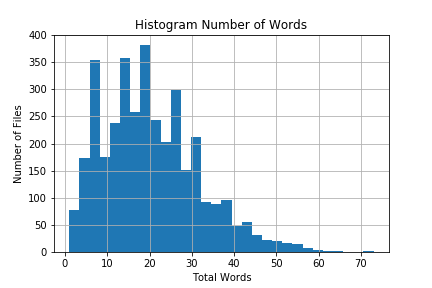} 
    \includegraphics[width=0.4\textwidth]{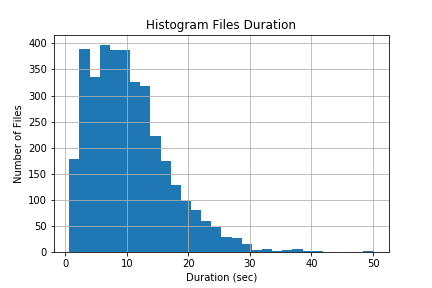} 
    \caption{The figure on the left shows the number of words and on the right the duration of each file.}
    \label{fig:dataset_histograms}
\end{figure*}

In the dataset, each audio file has its respective textual transcription (phonetic transcription is not provided). The final dataset consists of a total of $71,358$ words spoken by the speaker, $13,311$ unique words, resulting in $3,632$ audio files and totaling 10 hours and 28 minutes of speech. Audio files range in length from 0.67 to 50.08 seconds. The Figure \ref{fig:dataset_histograms} shows two histograms regarding the number of words and the duration of each file.

To compare TTS-Portuguese Corpus with datasets used in the literature for speech synthesis, we chose the LJ Speech \citep{ito2017lj} dataset, which is one of the most widely used, publicly available datasets, for training single-speaker models in the English language. Additionally, we present the statistics for the SSF1 dataset, which is a corpus of European Portuguese although not explored in the work of \cite{ttspropor}. Table \ref{fig:setencestypes} shows the language, duration, sampling rate and percentage of interrogative, exclamatory and declarative sentences in the LJ Speech, SSF1 and TTS-Portuguese datasets.




\begin{table*}[!htbp]
\caption{Comparison between LJ Speech, SSF1 and TTS-Portuguese datasets in terms of Language, Duration, Sampling Rate and proportion of sentences: Interrogative (Int), Exclamatory (Exc) and Declarative (Dec).}
\label{fig:setencestypes}
 \centering
\begin{tabular}{l|l|l|l|l|l|l}
\hline
\textbf{Dataset}        & \textbf{Language} & \textbf{Duration}& \textbf{Sampling rate} &\textbf{ Int.} & \textbf{Exc.} & \textbf{Dec. }  \\ \hline
LJ Speech    & EN & 24h &  22 kHz &0.58\%               & 0.35\%    & 99.07\%       \\ \hline
SSF1 & PT-PT& 14h  &  16 kHz &11\%  & 0.7\% & 88.3\% \\ \hline
This work & PT-BR &  10h & 48 kHz & 3.42\%  & 0.38\%      &   96.96\%     \\ \hline
\end{tabular}
\end{table*}

The TTS-Portuguese Corpus dataset has a smaller number of hours when compared to the others, it has 14 hours less than the LJ Speech and 4 hours less than the SSF1.

The sampling rate of 22 kHz is widely used in the training of TTS models based on deep learning. However, some works like \cite{tacotron2} use a sampling rate of 24 kHz. In addition, \cite{kumar2020nu} showed that it is possible to obtain a 44 kHz TTS model by training the NU-GAN model on a dataset sampled at 44 kHz. Following this idea, we made available the TTS-Portuguese Corpus at 48 kHz, a sampling rate much higher than the datasets compared here.
Finally, in the distribution of interrogative and exclamatory sentences, TTS-Portuguese Corpus has less coverage for these sentences compared to the SSF1 in-house dataset. However, TTS-Portuguese Corpus has greater coverage of these sentences than the publicly available dataset LJ Speech.

The TTS-Portuguese Corpus\footnote{Repository: https://github.com/Edresson/TTS-Portuguese-Corpus} is open source, and publicly available under the terms of the license Creative Commons Attribution 4.0 (CC BY 4.0)\footnote{https://creativecommons.org/licenses/by/4.0/}.
 

\color{black}
\section{Experiments} \label{sec:method}
To evaluate the quality of TTS-Portuguese Corpus in practice,  we explored the speech synthesis using models of prominence in the literature. We chose the models: DCTTS \citep{dctts} and  Tacotron~2 \citep{tacotron2}.

Here, we compare the models DCTTS and Tacotron 2.  To maintain results reproducible, we used open source implementations and tried to replicate related works as faithfully as possible. In the cases where hyper-parameters were not specified, we empirically optimized those for our dataset. 
We have used the following implementations: DCTTS provided by \cite{opendctts} and 
Tacotron 2 provided by \cite{ttsmozilla}. 

For all experiments, to speed up training, we initialized the model using the weights of the pre-trained model on English, using the LJ Speech dataset and we also use RTISI-LA \citep{zhu2007real} as a vocoder, which is a variation of the Griffin-Lim \citep{griffin1984signal} vocoder.


Although the acquisition avoided external noise as best as possible, the audio files were not recorded in a studio setting. Therefore, some noise may be present in part of the files. To reduce the interference with our analysis, we applied RNNoise \citep{valin2017hybrid} in all audio files. RNNoise is based on Recurrent Neural Networks; more specifically Gated Recurrent Unit \citep{cho2014learning}, and demonstrated good performance for noise suppression.

We report two experiments:
\begin{itemize}

\item \textbf{Experiment 1}: replicates the implementation of the DCTTS model, training the model for Portuguese with the TTS-Portuguese Corpus. For this experiment, as reported in the DCTTS article, the model receives the text directly as input, so no phonetic transcription is used. As previously mentioned, the original DCTTS paper does not describe any normalization, so for the model to converge we tested different normalization options and decided to use, in all layers, 5\% dropout and layer normalization \citep{ba2016layer}. We did not use a fixed learning rate as described in the original article. Instead, we used a starting learning rate of 0.001 decaying using Noam's learning rate decay scheme \citep{vaswani2017attention}.


\item \textbf{Experiment 2}: this experiment explores Tacotron 2 \citep{tacotron2} model, for that we use the Mozilla TTS implementation \citep{ttsmozilla}. This model receives phonetic transcription as input instead of text directly. To perform phonetic transcription we use the Phonemizer\footnote{https://github.com/bootphon/phonemizer} library that supports 121 languages and accents.
\end{itemize}

In experiment 1, two parts of the model are trained separately. The first part of the model, called Text2Mel, is responsible for generating a Mel spectrogram  from the input text and this part of the model was induced using the composition of the functions: binary cross-entropy, L1 \citep{deeplrgoodfellow} and guided attention loss \citep{dctts}. The second part, called SSRN, is responsible for the transformation of a mel spectrogram into the complete STFT spectrogram and applies super-resolution in the process and the loss function is composed of the functions L1 and binary cross-entropy.

In experiment 2, no guided attention is used, therefore, the loss function did not include the cost of attention. Since the network is trained end-to-end, the loss depends on the output of two network modules. The first module converts text into Mel spectrogram. The second module is a SSRN-like module called CBHG (1-D Convolution Bank Highway Network Bidirectional Gated Recurrent Unit). 

Table \ref{tab:computador} shows the hardware specifications of the equipment used for model training. Experiment 1 was trained on computer 2,  while experiment 2 were performed using computer 1.

\begin{table*}[!htbp]\small
\centering
\caption{Hardware specifications of the computers used in training.}
\label{tab:computador}
\begin{tabular}{l|l|l}
\hline
\textbf{Specifications}                                  & \textbf{Computer 1} &\textbf{ Computer 2}\\ \hline
Processor                          & i7-8700       &   i7-7700                       \\ \hline
RAM memory                         & 16 GB          & 32 GB                           \\ \hline
Video Card & Nvidia GeForce Gtx Titan V  &   Nvidia GeForce Gtx 1080 TI  \\ \hline
Operational system & Ubuntu 18.04 & Windows 10  \\ \hline
\end{tabular}
\end{table*} 

Table \ref{tab:dadosdotreinamento} presents the training data from the experiments. The metrics presented in the table are: number of training steps, and the time required for training. It is important to note that experiment 1 is trained in two phases, both reported in the table: Text2Mel and SSRN.

\begin{table*}[!htbp]\small
\centering
\caption{Model training.}
\label{tab:dadosdotreinamento}
\begin{tabular}{l|l|l}
\hline
\textbf{Experiment} & \textbf{Training steps}  & \textbf{Time} \\\hline 

Experiment 1 (Text2Mel/SSRN)  &       
2115k/2019k &       
4d19h/5d22h 
\\\hline 


Experiment 2 &
261k &    
9d7h
\\\hline

\end{tabular}
\end{table*}


\section{Results and Discussion} \label{sec:results}

To compare and analyze our results we used the Mean Opinion Score (MOS) calculated following the work of \cite{mos}. To calculate the MOS, the evaluators were asked to assess the naturalness of generated sentences on a five-point scale (1: Bad, 2: Poor, 3: Fair, 4: Good, 5: Excellent). We chose we chose 20 phonetically balanced sentences \citep{seara1994estudo} not seen in the training, so that our analysis has a good phonetic coverage. These sentences were synthesized for each of our experiments. In addition, 20 samples with the pronunciation of these sentences by the original speaker were added as ground truth. Each sample was evaluated by 20 native evaluators. Our Models, synthesized audios, corpus and an interactive demo are public available\footnote{https://edresson.github.io/TTS-Portuguese-Corpus/}.

\begin{figure}[!htbp]\small
 \centering
    \includegraphics[width=0.5\textwidth]{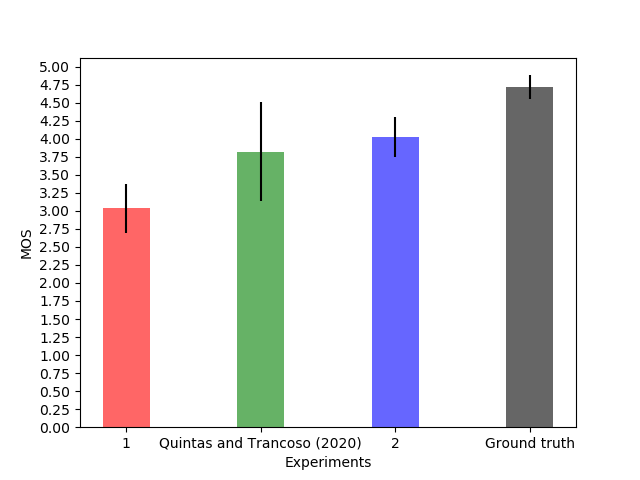} 
    \caption{MOS Analysis Chart.}
    \label{fig:graficorna}
\end{figure}

Table \ref{tab:resultados-mos}  presents the MOS values, with their respective 95\% confidence intervals, for our experiments and for the best experiment of the     \cite{ttspropor}, which can also be seen in Figure \ref{fig:graficorna}.


\begin{table}[!htbp]\small
\centering
\caption{MOS Results.}
\label{tab:resultados-mos}
   \begin{tabular}{c|c}
    \hline
    \textbf{Experiment}  & \textbf{MOS (Rank)} \\\hline
    Ground truth - SSF1   &  4.42  \textpm 0.60 (-)  \\\hline
    \cite{ttspropor} & 3.82 \textpm 0.69 (-) \\ \specialrule{1pt}{1pt}{1pt}
    Ground truth - Our   &  4.71 \textpm  0.16  (-)  \\\hline
                Experiment 1 &  3.03 \textpm  0.34 (2) \\\hline
                Experiment 2 &  \textbf{4.02} \textpm  \textbf{0.27} (1) \\\hline
    \end{tabular}
\end{table}
 
\color{black}
The results of the main analysis indicate that experiment 2 (Mozilla TTS) presented the best MOS value (4.02). 
According to \cite{mos}, the obtained value indicates a good quality audio, with a barely perceptible, but not annoying, distortion. On the other hand, experiment 1 presented a MOS of 3.03 indicating a perceptible and slightly annoying distortion in the audios. 

With respect to previous results in the English language, \cite{tacotron2} (Tacotron 2) can be compared to our experiment 2. The authors trained their model on an in-house US English dataset (24.6 hours), reaching a MOS of 4.52 \textpm 0.06. Considering the confidence intervals, our model reaches 4.29 in the best case and 3.75 in the worst case. Therefore, our model has a slightly lower MOS, and this can be justified as \cite{tacotron2} uses the WaveNet vocoder which achieves a higher quality in relation to the RTISI-LA/Griffin-Lim vocoder as shown in \citep{deepvoice3}.


Finally, regarding DCTTS we obtained 3.03 \textpm 0.34 MOS. The original paper \citep{dctts} had 2.71 \textpm 0.66 on the LJ Speech dataset. Therefore, the \cite{dctts} model can at best achieve a MOS of 3.37 and in the worst case 2.05. On the other hand, our model can at best achieve a MOS of 3.37 and in the worst case 2.69. Thus, the DCTTS model trained in the LJ Speech dataset and the TTS-Portuguese Corpus dataset showed similar MOS results.


It is also possible to compare our results with related works in Portuguese. The current state of the art (SOTA) in Portuguese \citep{ttspropor} achieved a MOS score of 3.82 \textpm 0.69 when training Tacotron 2 on the in-house SSF1 dataset. Considering the confidence intervals in the best case, the model of \cite{ttspropor} can achieve a MOS of 4.51 and in the worst case of 3.13. On the other hand, as previously discussed, our best model can reach 4.29 and 3.75 in the best and worst cases, respectively. These values are compatible since in the work of \cite{ttspropor} the authors used the neural vocoder WaveNet that generates speech with a higher quality. In addition, our confidence intervals are shorter, which may indicate that our evaluators agreed more during the evaluation. In addition, \cite{ttspropor} used only 8 evaluators in their MOS analysis, while in this work we used 20 evaluators; the number of evaluators can also have a impact on confidence intervals.

Comparing the Ground truth for the SSF1 dataset and TTS-Portuguese Corpus, we can see that the TTS-Portuguese Corpus can vary from 4.87 to 4.55 in the best and worst cases, respectively. On the other hand, the MOS for the SSF1 dataset reported by \cite{ttspropor} ranges from 5.02 (a value above 5 can be justified by rounding) to 3.82. Considering this MOS analysis, the two datasets are comparable in terms of quality and naturalness.
\color{black}

%



\section{Conclusions and Future Work} \label{sec:conc}

This work presented an open dataset, as well as the training of two speech synthesizer models based on deep learning, applied to the Brazilian Portuguese language. The dataset is publicly available and contains approximately 10.5 hours.

We found that it is possible to train a good quality speech synthesizer for Portuguese using our dataset, reaching 4.02 MOS value. Our best results were based on Tacotron 2 model. We had MOS scores comparable to the SOTA paper that explores the use of Deep Learning in the Portuguese language \citep{ttspropor}, using a in-house dataset. In addition, our results are also comparable to works in the literature that used the English language.


To the best of our knowledge, this is the first publicly available single-speaker synthesis dataset for the language. 
Similarly, the trained models are a contribution to the Portuguese language, since it has limited open access models based on deep learning.


In future work we intend to investigate the training of flow-based \citep{kingma2016improved, hoogeboom2019emerging, durkan2019neural} TTS models, such as Flowtron \citep{valle2020flowtron}, GlowTTS \citep{kim2020glow} and Flow-TTS \citep{miao2020flow}, in the TTS-Portuguese Corpus dataset.



\section{Acknowledgements}
This study was financed in part by the Coordena\c{c}\~{a}o de Aperfei\c{c}oamento de Pessoal de N\'{i}vel Superior -- Brasil (CAPES) -- Finance Code 001. 
We also gratefully acknowledge the support of NVIDIA corporation with the donation of the GPU used in part of the experiments presented in this research.





\FloatBarrier
\bibliographystyle{IEEEtran}
\bibliography{mybib}

\begin{thebibliography}{10}
\providecommand{\url}[1]{#1}
\csname url@samestyle\endcsname
\providecommand{\newblock}{\relax}
\providecommand{\bibinfo}[2]{#2}
\providecommand{\BIBentrySTDinterwordspacing}{\spaceskip=0pt\relax}
\providecommand{\BIBentryALTinterwordstretchfactor}{4}
\providecommand{\BIBentryALTinterwordspacing}{\spaceskip=\fontdimen2\font plus
\BIBentryALTinterwordstretchfactor\fontdimen3\font minus
  \fontdimen4\font\relax}
\providecommand{\BIBforeignlanguage}[2]{{%
\expandafter\ifx\csname l@#1\endcsname\relax
\typeout{** WARNING: IEEEtran.bst: No hyphenation pattern has been}%
\typeout{** loaded for the language `#1'. Using the pattern for}%
\typeout{** the default language instead.}%
\else
\language=\csname l@#1\endcsname
\fi
#2}}
\providecommand{\BIBdecl}{\relax}
\BIBdecl

\bibitem{purington2017alexa}
A.~Purington, J.~G. Taft, S.~Sannon, N.~N. Bazarova, and S.~H. Taylor, ``"
  alexa is my new bff" social roles, user satisfaction, and personification of
  the amazon echo,'' in \emph{Proceedings of the 2017 CHI Conference Extended
  Abstracts on Human Factors in Computing Systems}, 2017, pp. 2853--2859.

\bibitem{dempsey2017teardown}
P.~Dempsey, ``The teardown: Google home personal assistant,'' \emph{Engineering
  \& Technology}, vol.~12, no.~3, pp. 80--81, 2017.

\bibitem{siri}
T.~R. Gruber, ``Siri, a virtual personal assistant-bringing intelligence to the
  interface,'' in \emph{Semantic Technologies Conference}, 2009.

\bibitem{dctts}
H.~Tachibana, K.~Uenoyama, and S.~Aihara, ``Efficiently trainable
  text-to-speech system based on deep convolutional networks with guided
  attention,'' \emph{arXiv preprint arXiv:1710.08969}, 2017.

\bibitem{ze2013statistical}
H.~Ze, A.~Senior, and M.~Schuster, ``Statistical parametric speech synthesis
  using deep neural networks,'' in \emph{2013 ieee international conference on
  acoustics, speech and signal processing}.\hskip 1em plus 0.5em minus
  0.4em\relax IEEE, 2013, pp. 7962--7966.

\bibitem{teixeira2003prediction}
J.~P. Teixeira, D.~Freitas, and H.~Fujisaki, ``Prediction of fujisaki model's
  phrase commands,'' in \emph{Eighth European Conference on Speech
  Communication and Technology}, 2003.

\bibitem{klatt1980software}
D.~H. Klatt, ``Software for a cascade/parallel formant synthesizer,'' \emph{the
  Journal of the Acoustical Society of America}, vol.~67, no.~3, pp. 971--995,
  1980.

\bibitem{charpentier1986diphone}
F.~Charpentier and M.~Stella, ``Diphone synthesis using an overlap-add
  technique for speech waveforms concatenation,'' in \emph{ICASSP'86. IEEE
  International Conference on Acoustics, Speech, and Signal Processing},
  vol.~11.\hskip 1em plus 0.5em minus 0.4em\relax IEEE, 1986, pp. 2015--2018.

\bibitem{tokuda2000speech}
K.~Tokuda, T.~Yoshimura, T.~Masuko, T.~Kobayashi, and T.~Kitamura, ``Speech
  parameter generation algorithms for hmm-based speech synthesis,'' in
  \emph{2000 IEEE International Conference on Acoustics, Speech, and Signal
  Processing. Proceedings (Cat. No. 00CH37100)}, vol.~3.\hskip 1em plus 0.5em
  minus 0.4em\relax IEEE, 2000, pp. 1315--1318.

\bibitem{braude2013template}
D.~A. Braude, H.~Shimodaira, and A.~B. Youssef, ``Template-warping based speech
  driven head motion synthesis.'' in \emph{Interspeech}, 2013, pp. 2763--2767.

\bibitem{aroon2015statistical}
A.~Aroon and S.~Dhonde, ``Statistical parametric speech synthesis: A review,''
  in \emph{2015 IEEE 9th International Conference on Intelligent Systems and
  Control (ISCO)}.\hskip 1em plus 0.5em minus 0.4em\relax IEEE, 2015, pp. 1--5.

\bibitem{wang2011automatic}
W.~Y. Wang and K.~Georgila, ``Automatic detection of unnatural word-level
  segments in unit-selection speech synthesis,'' in \emph{2011 IEEE Workshop on
  Automatic Speech Recognition \& Understanding}.\hskip 1em plus 0.5em minus
  0.4em\relax IEEE, 2011, pp. 289--294.

\bibitem{siddhi2017survey}
D.~Siddhi, J.~M. Verghese, and D.~Bhavik, ``Survey on various methods of text
  to speech synthesis,'' \emph{International Journal of Computer Applications},
  vol. 165, no.~6, 2017.

\bibitem{deeplrgoodfellow}
I.~Goodfellow, Y.~Bengio, A.~Courville, and Y.~Bengio, \emph{Deep
  learning}.\hskip 1em plus 0.5em minus 0.4em\relax MIT press Cambridge, 2016,
  vol.~1.

\bibitem{kyle2017char2wav}
K.~K. J. F.~S. Kyle, K.~A. C. Y.~B. Jose, and S.~M. Sotelo, ``Char2wav:
  End-to-end speech synthesis,'' in \emph{International Conference on Learning
  Representations, workshop}, 2017.

\bibitem{tacotron}
Y.~Wang, R.~Skerry-Ryan, D.~Stanton, Y.~Wu, R.~J. Weiss, N.~Jaitly, Z.~Yang,
  Y.~Xiao, Z.~Chen, S.~Bengio \emph{et~al.}, ``Tacotron: A fully end-to-end
  text-to-speech synthesis model,'' \emph{arXiv preprint arXiv:1703.10135},
  2017.

\bibitem{tacotron2}
J.~Shen, R.~Pang, R.~J. Weiss, M.~Schuster, N.~Jaitly, Z.~Yang, Z.~Chen,
  Y.~Zhang, Y.~Wang, R.~Skerrv-Ryan \emph{et~al.}, ``Natural tts synthesis by
  conditioning wavenet on mel spectrogram predictions,'' in \emph{2018 IEEE
  International Conference on Acoustics, Speech and Signal Processing
  (ICASSP)}.\hskip 1em plus 0.5em minus 0.4em\relax IEEE, 2018, pp. 4779--4783.

\bibitem{deepvoice3}
W.~Ping, K.~Peng, A.~Gibiansky, S.~O. Arik, A.~Kannan, S.~Narang, J.~Raiman,
  and J.~Miller, ``Deep voice 3: 2000-speaker neural text-to-speech,''
  \emph{arXiv preprint arXiv:1710.07654}, 2017.

\bibitem{kim2020glow}
J.~Kim, S.~Kim, J.~Kong, and S.~Yoon, ``Glow-tts: A generative flow for
  text-to-speech via monotonic alignment search,'' \emph{arXiv preprint
  arXiv:2005.11129}, 2020.

\bibitem{valle2020flowtron}
R.~Valle, K.~Shih, R.~Prenger, and B.~Catanzaro, ``Flowtron: an autoregressive
  flow-based generative network for text-to-speech synthesis,'' \emph{arXiv
  preprint arXiv:2005.05957}, 2020.

\bibitem{teixeira2001phonetic}
J.~P. Teixeira, D.~Freitas, D.~Braga, M.~J. Barros, and V.~Latsch, ``Phonetic
  events from the labeling the european portuguese database for speech
  synthesis, feup/ipbdb,'' in \emph{Seventh European Conference on Speech
  Communication and Technology}, 2001.

\bibitem{alencar2008lsf}
V.~Alencar and A.~Alcaim, ``Lsf and lpc-derived features for large vocabulary
  distributed continuous speech recognition in brazilian portuguese,'' in
  \emph{2008 42nd Asilomar Conference on Signals, Systems and Computers}.\hskip
  1em plus 0.5em minus 0.4em\relax IEEE, 2008, pp. 1237--1241.

\bibitem{quintanilha2020open}
I.~M. Quintanilha, S.~L. Netto, and L.~W.~P. Biscainho, ``An open-source
  end-to-end asr system for brazilian portuguese using dnns built from newly
  assembled corpora,'' \emph{Journal of Communication and Information Systems},
  vol.~35, no.~1, pp. 230--242, 2020.

\bibitem{pratap2020mls}
V.~Pratap, Q.~Xu, A.~Sriram, G.~Synnaeve, and R.~Collobert, ``Mls: A
  large-scale multilingual dataset for speech research,'' \emph{Proc.
  Interspeech 2020}, pp. 2757--2761, 2020.

\bibitem{ito2017lj}
K.~Ito, ``The lj speech dataset,'' https://keithito.com/LJ-Speech-Dataset/,
  2017, accessed: 2020-04-29.

\bibitem{deepvoice}
S.~O. Arik, M.~Chrzanowski, A.~Coates, G.~Diamos, A.~Gibiansky, Y.~Kang, X.~Li,
  J.~Miller, J.~Raiman, S.~Sengupta \emph{et~al.}, ``Deep voice: Real-time
  neural text-to-speech,'' \emph{arXiv preprint arXiv:1702.07825}, 2017.

\bibitem{deepvoice2}
S.~O. Ar{\i}k, G.~Diamos, A.~Gibiansky, J.~Miller, K.~Peng, W.~Ping, J.~Raiman,
  and Y.~Zhou, ``Deep voice 2: Multi-speaker neural text-to-speech,''
  \emph{arXiv preprint arXiv:1705.08947}, 2017.

\bibitem{griffin1984signal}
D.~Griffin and J.~Lim, ``Signal estimation from modified short-time fourier
  transform,'' \emph{IEEE Transactions on Acoustics, Speech, and Signal
  Processing}, vol.~32, no.~2, pp. 236--243, 1984.

\bibitem{world}
M.~Morise, F.~Yokomori, and K.~Ozawa, ``World: a vocoder-based high-quality
  speech synthesis system for real-time applications,'' \emph{IEICE
  TRANSACTIONS on Information and Systems}, vol.~99, no.~7, pp. 1877--1884,
  2016.

\bibitem{WaveNet}
A.~Van Den~Oord, S.~Dieleman, H.~Zen, K.~Simonyan, O.~Vinyals, A.~Graves,
  N.~Kalchbrenner, A.~Senior, and K.~Kavukcuoglu, ``Wavenet: A generative model
  for raw audio,'' \emph{arXiv preprint arXiv:1609.03499}, 2016.

\bibitem{benesty2011speech}
J.~Benesty, J.~Chen, and E.~A. Habets, \emph{Speech enhancement in the STFT
  domain}.\hskip 1em plus 0.5em minus 0.4em\relax Springer Science \& Business
  Media, 2011.

\bibitem{yu2015multi}
F.~Yu and V.~Koltun, ``Multi-scale context aggregation by dilated
  convolutions,'' \emph{arXiv preprint arXiv:1511.07122}, 2015.

\bibitem{kalchbrenner2016neural}
N.~Kalchbrenner, L.~Espeholt, K.~Simonyan, A.~v.~d. Oord, A.~Graves, and
  K.~Kavukcuoglu, ``Neural machine translation in linear time,'' \emph{arXiv
  preprint arXiv:1610.10099}, 2016.

\bibitem{zhu2007real}
X.~Zhu, G.~T. Beauregard, and L.~L. Wyse, ``Real-time signal estimation from
  modified short-time fourier transform magnitude spectra,'' \emph{IEEE
  Transactions on Audio, Speech, and Language Processing}, vol.~15, no.~5, pp.
  1645--1653, 2007.

\bibitem{bahdanau2014neural}
D.~Bahdanau, K.~Cho, and Y.~Bengio, ``Neural machine translation by jointly
  learning to align and translate,'' \emph{arXiv preprint arXiv:1409.0473},
  2014.

\bibitem{srivastava2015highway}
R.~K. Srivastava, K.~Greff, and J.~Schmidhuber, ``Training very deep
  networks,'' in \emph{Advances in neural information processing systems},
  2015, pp. 2377--2385.

\bibitem{chung2014empirical}
J.~Chung, C.~Gulcehre, K.~Cho, and Y.~Bengio, ``Empirical evaluation of gated
  recurrent neural networks on sequence modeling,'' \emph{arXiv preprint
  arXiv:1412.3555}, 2014.

\bibitem{wavenetmodificado}
A.~Tamamori, T.~Hayashi, K.~Kobayashi, K.~Takeda, and T.~Toda,
  ``Speaker-dependent wavenet vocoder,'' in \emph{Proceedings of Interspeech},
  2017, pp. 1118--1122.

\bibitem{ttspropor}
S.~Quintas and I.~Trancoso, ``Evaluation of deep learning approaches to
  text-to-speech systems for european portuguese,'' in \emph{International
  Conference on Computational Processing of the Portuguese Language}.\hskip 1em
  plus 0.5em minus 0.4em\relax Springer, 2020, pp. 34--42.

\bibitem{seara1994estudo}
I.~Seara, ``Estudo estat{\'\i}stico dos fonemas do portugu{\^e}s brasileiro
  falado na capital de santa catarina para elabora{\c{c}}{\~a}o de frases
  foneticamente balanceadas,'' Ph.D. dissertation, Disserta{\c{c}}{\~a}o de
  Mestrado, Universidade Federal de Santa Catarina~…, 1994.

\bibitem{kumar2020nu}
R.~Kumar, K.~Kumar, V.~Anand, Y.~Bengio, and A.~Courville, ``Nu-gan: High
  resolution neural upsampling with gan,'' \emph{arXiv preprint
  arXiv:2010.11362}, 2020.

\bibitem{opendctts}
K.~Park, ``A tensorflow implementation of dc-tts,''
  https://github.com/kyubyong/dc\_tts, 2018.

\bibitem{ttsmozilla}
E.~Gölge, ``Deep learning for text to speech,''
  https://github.com/mozilla/TTS, 2019.

\bibitem{valin2017hybrid}
J.-M. Valin, ``A hybrid dsp/deep learning approach to real-time full-band
  speech enhancement,'' \emph{arXiv preprint arXiv:1709.08243}, 2017.

\bibitem{cho2014learning}
K.~Cho, B.~Van~Merri{\"e}nboer, C.~Gulcehre, D.~Bahdanau, F.~Bougares,
  H.~Schwenk, and Y.~Bengio, ``Learning phrase representations using rnn
  encoder-decoder for statistical machine translation,'' \emph{arXiv preprint
  arXiv:1406.1078}, 2014.

\bibitem{ba2016layer}
J.~L. Ba, J.~R. Kiros, and G.~E. Hinton, ``Layer normalization,'' \emph{arXiv
  preprint arXiv:1607.06450}, 2016.

\bibitem{vaswani2017attention}
A.~Vaswani, N.~Shazeer, N.~Parmar, J.~Uszkoreit, L.~Jones, A.~N. Gomez,
  {\L}.~Kaiser, and I.~Polosukhin, ``Attention is all you need,'' in
  \emph{Advances in Neural Information Processing Systems}, 2017, pp.
  5998--6008.

\bibitem{mos}
F.~Ribeiro, D.~Flor{\^e}ncio, C.~Zhang, and M.~Seltzer, ``Crowdmos: An approach
  for crowdsourcing mean opinion score studies,'' in \emph{Acoustics, Speech
  and Signal Processing (ICASSP), 2011 IEEE International Conference on}.\hskip
  1em plus 0.5em minus 0.4em\relax IEEE, 2011, pp. 2416--2419.

\bibitem{kingma2016improved}
D.~P. Kingma, T.~Salimans, R.~Jozefowicz, X.~Chen, I.~Sutskever, and
  M.~Welling, ``Improved variational inference with inverse autoregressive
  flow,'' in \emph{Advances in neural information processing systems}, 2016,
  pp. 4743--4751.

\bibitem{hoogeboom2019emerging}
E.~Hoogeboom, R.~v.~d. Berg, and M.~Welling, ``Emerging convolutions for
  generative normalizing flows,'' \emph{arXiv preprint arXiv:1901.11137}, 2019.

\bibitem{durkan2019neural}
C.~Durkan, A.~Bekasov, I.~Murray, and G.~Papamakarios, ``Neural spline flows,''
  in \emph{Advances in Neural Information Processing Systems}, 2019, pp.
  7511--7522.

\bibitem{miao2020flow}
C.~Miao, S.~Liang, M.~Chen, J.~Ma, S.~Wang, and J.~Xiao, ``Flow-tts: A
  non-autoregressive network for text to speech based on flow,'' in
  \emph{ICASSP 2020-2020 IEEE International Conference on Acoustics, Speech and
  Signal Processing (ICASSP)}.\hskip 1em plus 0.5em minus 0.4em\relax IEEE,
  2020, pp. 7209--7213.

\end{thebibliography}
\end{document}